\documentclass[preprint,prl,superscriptaddress,prbib,11pt]{revtex4-1}
\usepackage{graphicx}
\usepackage{bm}
\usepackage{hyperref}
\usepackage{todonotes}
\usepackage{nicefrac}
\usepackage{bbm,braket}
\usepackage{bbm}
\usepackage{amsmath}
\usepackage[normalem]{ulem}
\usepackage{booktabs}
\usepackage{subfigure}
\usepackage{rotating}
\usepackage{ragged2e}
\usepackage{color}
\usepackage{float}
\usepackage[T1]{fontenc}
\usepackage[utf8]{inputenc}
\patchcmd{\subsubsection}{\itshape}{\bfseries}{}{}
\usepackage{caption}
\usepackage{soul}

\newcommand{\bk}{\bm k}

\newcommand{\bu}{\bm u}

\newcommand{\LM}{L_M}

\usepackage{geometry}
\begin{document}

\title{The Flexoelectric Twist: Persistence of Heavy Fermion Picture in Small-Angle Twisted Bilayer Graphene}

\author{Qirong Yao}
\thanks{These authors contributed equally.}
\affiliation{Department of Physics, The Hong Kong University of Science and Technology, Clear Water Bay, Kowloon, Hong Kong SAR}

\author{Ka Cheuk Fan}
\thanks{These authors contributed equally.}
\affiliation{Department of Physics, The Hong Kong University of Science and Technology, Clear Water Bay, Kowloon, Hong Kong SAR}

\author{Tianyu Qiao}
\thanks{These authors contributed equally.}
\affiliation{Department of Physics, The Hong Kong University of Science and Technology, Clear Water Bay, Kowloon, Hong Kong SAR}

\author{Hao Shi}
\affiliation{Department of Physics, The Hong Kong University of Science and Technology, Clear Water Bay, Kowloon, Hong Kong SAR}

\author{Jiangchang Zheng}
\affiliation{Department of Physics, The Hong Kong University of Science and Technology, Clear Water Bay, Kowloon, Hong Kong SAR}

\author{Kenji Watanabe}
\affiliation{National Institute for Materials Science, Tsukuba, Japan}

\author{Takashi Taniguchi}
\affiliation{National Institute for Materials Science, Tsukuba, Japan}

\author{Xi Dai}
\email{daix@ust.hk}
\affiliation{Department of Physics, The Hong Kong University of Science and Technology, Clear Water Bay, Kowloon, Hong Kong SAR}

\author{Berthold J\"ack}
\email{bjaeck@ust.hk}
\affiliation{Department of Physics, The Hong Kong University of Science and Technology, Clear Water Bay, Kowloon, Hong Kong SAR}

\date{\today}

\begin{abstract}
Twisted bilayer graphene (TBG) near the first ``magic angle'' ($\theta\approx1.05^{\circ}$) hosts topological flat bands that provide a fertile ground for exploring strongly correlated phases. At smaller twist angles ($\theta<1^{\circ}$), strong lattice relaxation profoundly alters the moiré landscape, but its impact on the electronic structure and appearance of other predicted magic angles through electro-mechanical coupling remains unresolved. Using scanning tunneling microscopy/spectroscopy on small-angle TBG devices with different twist angles $\theta\approx0.3-0.7^{\circ}$, we observe a pronounced $dI/dV$ peak at the AA site of the moiré lattice on top of a spectrum that is asymmetric about the charge neutrality point; we visualize lattice reconstruction at moiré lattice length scales (20-45\,nm); and we detect a dilation of the atomic lattice at the AA sites at Angstrom length scale. We demonstrate at a phenomenological level that the inclusion of moiré lattice reconstruction supports compact localized heavy fermion $f-$states at AA sites. The additional inclusion of atomic lattice dilation generates a strong flexo-electric response that explicitly breaks particle-hole symmetry, pushing the $f$-states down in energy. Consequently, these heavy fermion states, which are eightfold degenerate, energetically coexist with itinerant $c$-electrons at the AB/BA sites. Our spatially resolved spectroscopic STM measurements quantitatively visualize this coexistence, confirming the persistence of the heavy fermion model over a continuous twist-angle range at $\theta<1^{\circ}$. Our findings show that flexo-electric responses can realize qualitatively distinct electronic landscapes in moiré materials beyond the single-particle picture. Looking ahead, small-angle TBG could prove an ideal platform to explore the interacting large-$N$ Kondo lattice problem with high tunability of the $f-$occupation and $c-$density through twist-angle and electrostatic back gates, respectively.
\end{abstract}

\maketitle

\section{Introduction}

Moiré superlattices composed of two or more layers of rotationally misaligned two-dimensional materials have emerged as a premier platform for examining topological many-body states~\cite{andrei2021marvels, mak2022semiconductor}. In twisted bilayer graphene (TBG) near the first magic angle ($\theta\approx1.05^\circ$), the moiré potential modulation of the original Dirac electrons generates a pair of nearly particle-hole (PH) symmetric flat bands~\cite{lopes2007graphene, li2010observation, bistritzerMoireBandsTwisted2011, andrei2020graphene}. These flat bands (spectral width $\approx20\,$meV~\cite{lisi2021observation}) provide a fertile ground for exploring a rich phase diagram that hosts exotic phenomena, including superconductivity, correlated insulators, ferromagnetism and integer and fractional Chern insulators~\cite{cao2018unconventional, cao2018correlated, sharpe2019emergent, nuckolls2020strongly, xie2021fractional}.

\begin{figure}[h]
    \centering
    \includegraphics[width=\textwidth]{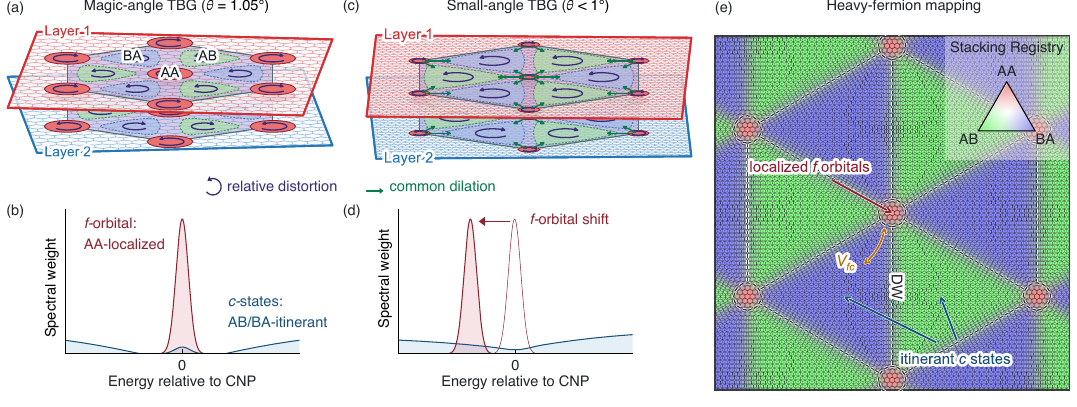}
    \caption{\justifying{\bf The flexoelectric twist--electronic structure of small-angle twisted bilayer graphene:} (a,c) Schematic lattice reconstruction near the first magic angle and at smaller twist angles, respectively. Opposite displacements of the two layers enlarge the AB/BA domains relative to the AA regions. In (c), a common in-plane dilation (green arrows) additionally expands the atomic lattice near AA. Curved arrows denote relative distortion; Layer rotations are exaggerated for clarity. 
    (b,d) Schematic spectral distributions near the first magic angle and at small twist angles, respectively. Red and blue curves indicate AA-localized and itinerant spectral weight. Near the first magic angle, the hybridized bands form an isolated flat-band manifold with approximate particle--hole (PH) symmetry about the charge neutrality point (CNP). At small angles, AA-localized resonances persist within a dense itinerant continuum. The scalar deformation potential associated with common dilation shifts the dominant AA resonance below CNP.
    (e) Real-space heavy-fermion representation common to both regimes. The AA-centered $f-$orbitals couple through $V_{fc}$ to the itinerant $c-$states, which have substantial weight in the AB/BA domains and domain-wall (DW) network. The color map denotes the local stacking registry.}
    \label{fig:fig1}
\end{figure}

It has recently been recognized that the effective low-energy electronic structure of magic-angle TBG can be elegantly captured by a topological heavy fermion (HF) model~\cite{shiHeavyFermionRepresentation2022, SongBernevig2022}. This model partitions the system into $f$-orbitals localized at the AA stacking sites and itinerant $c$-electrons residing in the neighboring AB- and BA-stacking regions [Fig.~\ref{fig:fig1}(a)]. At the first magic angle, owing to the negligible density of states of the $c$-electrons near the PH-symmetric flat bands [Fig.~\ref{fig:fig1}(b)], the phenomenology of the partially filled flat bands is dominated by on-site Coulomb interactions within the $f$-orbitals, whereas the effect of coupling to the $c$-states is largely negligible except within a small region around the moir\'e $\Gamma$ point~\cite{SongBernevig2022, shiHeavyFermionRepresentation2022}.

However, as the twist angle decreases to $\theta<1^{\circ}$, lattice relaxation of the metastable moiré landscape becomes increasingly dominant. This results in a relative reduction of the AA stacking area and extended AB/BA domains with Bernal stacking registry separated by domain wall states [Fig.~\ref{fig:fig1}(c)]~\cite{alden2013strain, san2013helical, zhang2013valley}. Strain gradients cause a flexoelectric response~\cite{mcgilly2020visualization, li2021unraveling}, which is expected to redistribute electric charge within the moiré unit cell. Even though continuum models predicted that other magic angles should occur at $\theta<1^{\circ}$~\cite{bistritzerMoireBandsTwisted2011}, the impact of this lattice reconstruction on the low-energy electronic structure through electro-mechanical coupling has remained poorly understood. Consequently, the experimental study of TBG at these smaller twist angles has received comparatively less attention~\cite{alden2013strain, huang2018topologically, rickhaus2018transport, xu2019giant, nuckolls2025spectroscopy}.

Here, using scanning tunneling microscopy and spectroscopy (STM/STS) experiments on small twist-angle TBG devices ranging between $\theta\approx0.3^\circ$ and $1.1^\circ$, we directly examine the effect of lattice reconstruction on the low-energy electronic structure. We detect enlarged AB/BA at moiré length scales as well as an atomic lattice dilation at the AA sites, accompanied by a highly electron-hole asymmetric differential tunnel conductance ($dI/dV$) spectrum with a pronounced peak localized at the AA site. By comparing these observations with quantitative model calculations, we reveal that the inclusion of this lattice dilation [green arrows in Fig.~\ref{fig:fig1}(c)] introduces a scalar strain field that naturally breaks electron-hole symmetry at small twist angles [Fig.~\ref{fig:fig1}(d)].

The resulting electronic landscape provides a direct mapping to the HF model~\cite{shiHeavyFermionRepresentation2022, SongBernevig2022} of highly localized $f$-orbitals highly localized $f$ orbitals embedded in a broad, interconnected continuum of itinerant $c$ states. [Fig.~\ref{fig:fig1}(e)]. More broadly, our work reveals that the low-energy physics of small-angle TBG can be captured by the HF model over a remarkably large angle range without the need for fine-tuning to a ``magic'' condition. This establishes small-angle TBG as a highly tunable framework for examining the Kondo lattice phase diagram and exploring complex strongly correlated states in the future.

\section{Electron-hole symmetry breaking and lattice dilation detected with Scanning Tunneling Microscopy}

\begin{figure}[h]
    \centering
    \includegraphics[width=\textwidth]{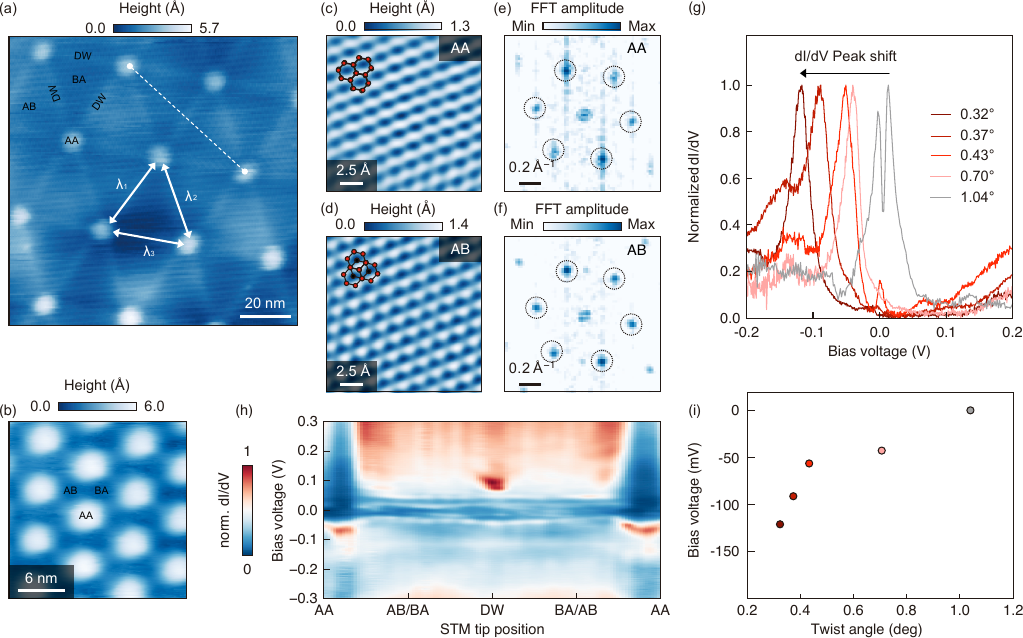}
    \caption{\justifying{\bf Visualizing the impact of moiré lattice relaxation on the electronic properties of twisted bilayer graphene:} (a) STM topography of small-angle TBG sample with twist angle $\theta \approx$  0.37\textsuperscript{$\circ$} (bias voltage $V_{B}$ = -0.6\,V, tunnel current $I=$50\,pA). (b) STM topography of a TBG sample ($\theta \approx$  1.16\textsuperscript{$\circ$}) near the first magic angle ($V_{B}=-0.3$\,V, $I=1\,$nA). (c,~d) High-resolution STM topographies recorded at the center of an AA and AB site, respectively, within the field of view of panel a ($V_{B} = 0.3\,$V, $I=50\,$pA). (e,~f) Corresponding two-dimensional fast Fourier transform (FFT) maps of panels c and d.  (g) Normalized $dI/dV$ spectra recorded at the center of AA sites of TBG samples with different indicated twist angles ($V_{\rm B}=-300\,$mV, $I=500\,$pA, $V_{\rm mod}=3\,$mV). (h) Spectroscopic line cut recorded along the AA-AB/BA-DW-BA/AB-AA direction within the field of view of panel a ($V_{\rm B}=-300\,$mV, $I = 500\,$pA, lock-in modulation $V_{\rm mod}=3\,$mV). (i) Twist angle dependence of the bias voltage position of the $dI/dV$ peak detected on AA sites in panel g.}
    \label{fig:fig2}
\end{figure}

We have used cryogenic (temperature $T=4.5\,$K) scanning tunneling microscopy (STM) measurements to characterize the impact of lattice relaxation on the topographic and electronic properties of twisted bilayer graphene (TBG). To this end, we have studied TBG devices near the first magic angle ($\theta\approx1.1^{\circ}$) and at smaller twist angles ($\theta\approx0.3^{\circ}-0.7^{\circ}$). We determined $\theta$ and small hetero-strain $\epsilon$, which is present in all devices, by fitting a uniaxial heterostrain model~\cite{kerelsky2019maximized} using the experimentally determined moiré lattice wavelengths $\lambda_{1}, \lambda_{2}, \lambda_{3}$ [see Fig.~\ref{fig:fig2}(a)] as input parameters. Details of the TBG device fabrication, STM experiment, and this fitting procedure are described in the Methods section and Supplementary Section~1.

In Fig.~\ref{fig:fig2}(a,~b), we compare STM topographies recorded on small-angle TBG ($\theta\approx0.37^\circ$) and on a sample near the magic angle ($\theta\approx1.17^{\circ}$). Both topographies exhibit a triangular lattice of AA stacking sites, but the different twist angle strongly affects the moiré lattice characteristics. At small angles, the AA sites are separated by much larger, extended AB stacking regions and connected by domain wall sites [also see previous work on the STM study of the AB and domain wall sites by Huang {\em et al.} in Ref.~\cite{huang2018topologically}]. This finding contrasts with the dense lattice of AA sites that characterizes the well-established topography of magic-angle devices~\cite{xie2019spectroscopic, kerelsky2019maximized, choi2019electronic}, highlighting the profound role of lattice relaxation on the moiré lattice at small twist angles $\theta<1^\circ$. 

Indications for lattice relaxation effects can also be found on atomic length scales when examining atomically resolved STM topographies recorded at the AA and AB sites of the small-angle TBG sample ($\theta\approx0.37^\circ$) in Fig.~\ref{fig:fig2}(c,~d). Note that the different topographic appearances of these maps reflect the underlying atomic stacking registry of the AA and AB regions. The atomic lattice constants of these regions ($a_{\rm AA}$ and $a_{\rm AB}$) can be precisely determined by analyzing the corresponding two-dimensional fast-Fourier transform (2D-FFT) maps shown in Fig.~\ref{fig:fig2}(e,~f) in which the periodic atomic lattice in real space transforms into a set of six Bragg vectors ${\bf q}_{\rm Bragg}$ of length $q_{\rm Bragg}=2/\sqrt{3}a_{\rm AA/AB}$. Our detailed analysis of the FFT maps [see Supplementary Section~2] reveals a relative dilation $\Delta\eta=100\times2(a_{\rm AA}-a_{\rm AB})/(a_{\rm AA}+a_{\rm AB})\approx2.1\,\%$ of the atomic lattice constant between the AA and AB sites averaged over the field of view of Fig.~\ref{fig:fig2}(a); that is, the atomic lattice constant of the AA site is larger compared to the lattice constant of the AB site. In contrast, the same analysis applied to STM topographies recorded on the TBG device near the magic angle reveals identical atomic lattice constants at the AA and AB/BA sites [see Supplementary Section~2].

Moiré lattice relaxation effects also manifest in the local electronic density of states (LDOS) at low energies, indicating the presence of finite electro-mechanical coupling. Scanning tunneling spectroscopy (STS) measurements of a device near the magic angle ($\theta=1.04^{\circ}$) reveal a pair of $dI/dV$ peaks that is approximately symmetric about the charge neutrality point at zero bias voltage $V_{\rm B}$. These $dI/dV$ peaks, which give rise to the topographic contrast of the AA sites in Fig.~\ref{fig:fig2}(b), reflect the LDOS of the two flat bands~\cite{li2010observation, xie2019spectroscopic, kerelsky2019maximized, choi2019electronic} that preserve approximate PH symmetry~\cite{Bernevig2021TBGIII}. STS measurements conducted on small-angle TBG ($\theta\approx0.37^\circ$) reveal qualitatively different characteristics. The $dI/dV$ spectrum [Fig.~\ref{fig:fig2}(g)] features a pronounced peak centered at $V_{\rm B}\approx-80\,$mV and exhibits a pronounced asymmetry about the charge neutrality point (CNP), which is located at zero bias voltage. A spectroscopic line-cut recorded between two AA sites on the sample surface [white dashed line in Fig.~\ref{fig:fig2}(a)] illustrates that this $dI/dV$ peak is strongly localized to the AA site [Fig.~\ref{fig:fig2}(h)]. Moreover, it reveals a $dI/dV$ peak at $V_{\rm B}\approx80\,$mV localized at the domain wall site, which was previously discussed in the context of helical edge states~\cite{huang2018topologically}, as well as a set of spatially extended $dI/dV$ features of smaller amplitude between $-50\,\text{mV}<V_{\rm B}<50\,\text{mV}$ that appear as sharp spectroscopic features of width $\approx10\,$mV. Finally, examining $dI/dV$ spectra recorded at the AA site of samples with different twist angles [Fig.~\ref{fig:fig2}(g)], we find that the position of the $dI/dV$ peak shifts to more negative bias voltages as the twist angle decreases [Fig.~\ref{fig:fig2}(i)].

\section{Impact of axial strain on the low-energy electronic structure in small-angle TBG}

Our experimental observations of lattice reconstruction at moiré length scales, relative atomic lattice dilation at AA sites, and a particle--hole-asymmetric local density of states suggest an important role of lattice deformation effects in small-angle TBG at $\theta<1$°. To interpret these observations, we distinguish the relative reconstruction at moiré length scales from atomic-scale dilation effects. From a theoretical viewpoint, toward smaller twist angles, lattice relaxation effects cause the two graphene layers to rotate locally toward Bernal registry over larger AB/BA domains, concentrating the registry changes near the AA centers and a well-developed domain-wall network [Fig.~\ref{fig:fig1}(a,~c)]~\cite{NamKoshino2017,Carr2018Relaxation,kazmierczakStrainFieldsTwisted2021,WaletGuinea2019MarginalChannels,huang2018topologically}. We obtain this relative reconstruction by minimizing the continuum elastic and interlayer adhesion energy [see Supplementary Section~3]. However, within the effective continuum model used here, relative reconstruction preserves the approximate particle--hole (PH) symmetry of the ZLL-derived spectrum [see Ref.~\cite{Bernevig2021TBGIII} and Supplementary Section~4]. Microscopic calculations near the first magic angle report PH-breaking corrections of several meV~\cite{KangVafek2022PH}, much smaller than the energy scale corresponding to the approximately $100\,\mathrm{mV}$ shift of the experimentally observed $dI/dV$ peak in small-angle devices.


Motivated by our experimental observations, we therefore introduce an additional phenomenological common displacement that expands the atomic lattice of both graphene layers near the AA site center relative to the AB/BA regions. We show that the inclusion of this dilation, which effectively introduces a finite electro-mechanical coupling, naturally results in a PH symmetry-breaking LDOS through the scalar deformation potential [Fig.~\ref{fig:fig1}(c)]. We assume equal additional displacements in the two layers, i.e., a common displacement that leaves the relative stacking registry unchanged, which gives a lattice constant dilation profile $\eta(\bm r) = \frac{2\Delta\eta}{9}[C_p(\bm r) - \braket{C_p}_M]$, with $C_p(\bm r) = \sum_{j=1}^3 \cos[\bm b_j\cdot\bm\delta_{\rm ref}(\bm r)]$. Here, $\Delta\eta = \eta_{\rm AA} - \eta_{\rm AB}$ denotes the lattice dilation contrast between AA and AB sites, and $\braket{\cdot}_M$ denotes the moir\'e-cell average. The vectors $\bm b_j$ ($j=1,2,3$) are chosen from the first shell of the graphene reciprocal lattice and are separated by $120^\circ$. The reference stacking registry field $\bm\delta_{\rm ref}(\bm r)$ denotes the in-plane relative shift between two layers in a reference TBG model, whose twist angle can be tuned to obtain different expansion region sizes centered at AA [see Supplementary Section~3]. Through the scalar deformation potential of graphene~\cite{KangVafek2022PH, vafekContinuumEffectiveHamiltonian2023, ChoiJhiSon2010Strain, Grassano2020WorkFunction, Wang2021GlobalScalar}, this common dilation lowers the orbital energy in the AA stacking region and shifts the dominant ZLL-derived resonance below the CNP.

\begin{figure}[h]
    \centering
    \includegraphics[width=1\linewidth]{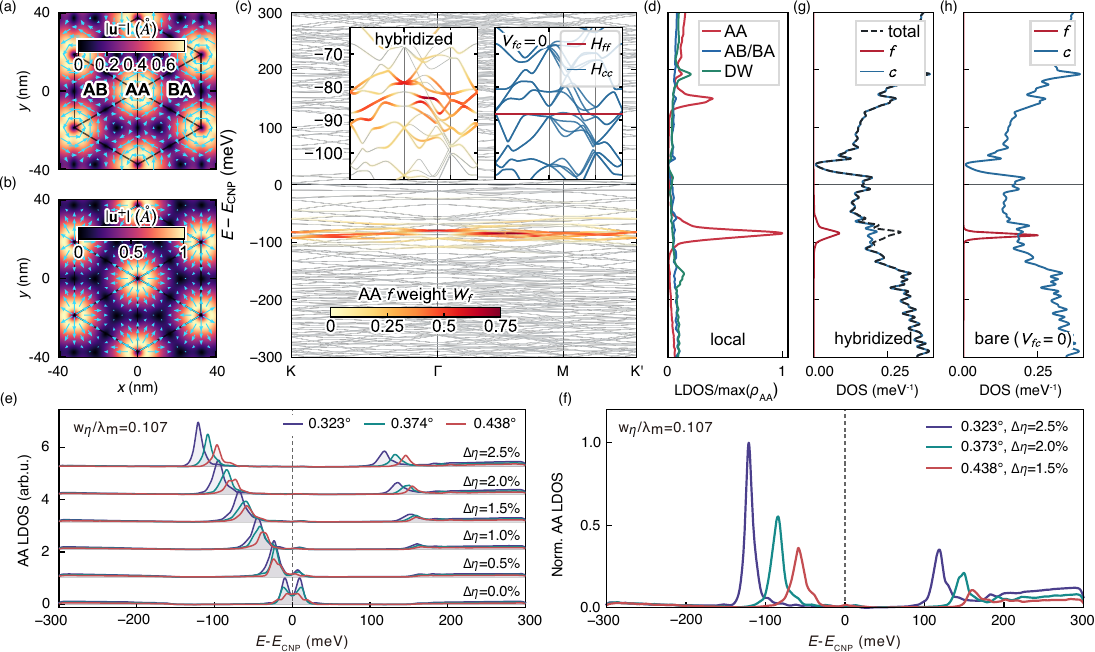}
     \caption{\justifying{\bf Particle-hole symmetry breaking from AA-site lattice dilation in small-angle twisted bilayer graphene:} 
     (a,~b) Relative and common displacement fields, $\bu^-=\bu^{(2)}-\bu^{(1)}$ and $\bu^+=\bu^{(2)}+\bu^{(1)}$, respectively, at $\theta=0.374^\circ$. The relative displacement is obtained from elastic relaxation, and the common displacement is introduced phenomenologically with an AA--AB dilation contrast $\Delta\eta=2\%$ and a half-contrast equivalent-area radius $w_\eta=0.107 \lambda_m$.
     (c) Continuum band structure colored by the AA-localized $f$-orbital weight. The left inset provides a zoomed-in view of the hybridized resonance, while the right inset shows the bare $f-$ and $c-$bands.
     (d) Registry-resolved local density of states (LDOS) at the AA, AB/BA, and DW sites.
     (e) Calculated AA site LDOS at three indicated twist angles as the lattice dilation contrast $\Delta\eta$ increases from $0$ to $2.5\%$. Spectra are vertically offset for clarity.
     (f) Representative calculated LDOS spectrum at the AA site at $\theta=0.323^\circ$, $0.374^\circ$, and $0.438^\circ$, using $\Delta\eta=2.5\%$, $2\%$, and $1.5\%$, respectively. Both (e) and (f) use $w_\eta/\lambda_m=0.107$, with vertical dashed lines marking the CNP.
     (g,~h) Hybridized total and $f/c-$projected DOS, and bare $f/c$ DOS at $V_{fc}=0$, respectively.}
    \label{fig:fig3}
\end{figure}

Figure~\ref{fig:fig3} presents continuum calculations that incorporate lattice relaxation and strain-dependent electronic couplings~\cite{NamKoshino2017, Carr2018Relaxation, KoshinoNam2020Continuum, KangVafek2022PH, vafekContinuumEffectiveHamiltonian2023}.
Details of the model, displacement-field construction, orbital projectors, and numerical checks are presented in Supplementary Section~4.
Figures~\ref{fig:fig3}(a,b) show the relative and common displacement fields at $\theta=0.374^\circ$, respectively.
The relative displacement produces opposite local rotations in the two layers, enlarging the AB/BA domains and concentrating the
distortion near the DW network. The associated strain generates a spatially modulated pseudovector
potential~\cite{KoshinoNam2020Continuum,KangVafek2022PH,vafekContinuumEffectiveHamiltonian2023,xieLatticeDistortionsMoire2023,Shi2019PMF,ceferinoPseudomagneticFieldsFully2024}, which by itself, however, \textit{does not} shift the DOS peak of $f$-orbitals localized at the AA sites.

The calculated spectral function [Fig.~\ref{fig:fig3}(c)] reveals a pronounced resonance near
$-85\,\mathrm{meV}$ within a dense continuum of weakly dispersive [band width $\approx10\,$meV] mini bands that occupy the entire displayed energy interval. These mini bands arise from band folding into the small moir\'e Brillouin zone and energetically overlap, collectively forming a broad itinerant background. Their narrow individual bandwidths therefore do not imply the same localization as the nearly dispersionless bare $f$-orbital bands~\cite{shiHeavyFermionRepresentation2022}. 
To identify the AA-localized component across the hybridized bands, we use the most localized zero-mode doublet of the continuum Hamiltonian expanded near an AA center [see Supplementary Section~5]. This local Hamiltonian has a pseudo-Landau-level form, providing AA-centered orbital projectors and a common construction for comparison with magic-angle TBG~\cite{shiHeavyFermionRepresentation2022}. 
The orbital character of this resonance is resolved using the AA-site localized zeroth Landau level (ZLL) doublet by the projector onto the primary $m=0,1$ ZLL doublet, comprising two orbitals per spin and valley and defining the $f$ sector of the heavy-fermion representation~\cite{SongBernevig2022, shiHeavyFermionRepresentation2022}. The band coloring resolves this doublet's weight across several hybridized bands near the resonance [left inset of Fig.~\ref{fig:fig3}(c)]. Examining the registry-resolved LDOS [Fig.~\ref{fig:fig3}(d)] confirms that the resonance at $E\approx-86\,$meV is strongly localized at the AA site, displaying much weaker amplitude at AB/BA and DW regions. 

To further verify the impact of lattice dilation on the low-energy electronic structure, we analyze the LDOS at the AA site as a function of the lattice dilation contrast $\Delta\eta$ [Fig.~\ref{fig:fig3}(e)]. In the absence of lattice dilation effects, the resulting LDOS is centered about the CNP and has approximate PH-symmetry at all twist angles. This case contrasts with the PH-breaking LDOS in the presence of lattice dilation ($\Delta\eta>0$), which is asymmetric about the CNP. Here, the dilation effect causes a twist-angle dependent shift of the AA resonance peak to lower energies. This finding highlights the critical role of common lattice dilation at the AA site, as detected in STM measurements, for inducing a PH-symmetry breaking band structure.

Finally, we compare the calculated LDOS at the AA site with our experimental observations from STM measurements [Fig.~\ref{fig:fig2}(g)]. In Fig.~\ref{fig:fig3}(f), we compare calculated LDOS spectra at the AA site for $\theta=0.323^\circ$, $0.374^\circ$, and $0.438^\circ$, using $\Delta\eta=2.5\%$, $2\%$, and $1.5\%$, respectively, at the common fractional width $w_\eta/\LM=0.107$. At each angle, the lattice dilation contrast $\Delta\eta$ was selected within the estimated strain range to match the measured energy $E=eV_{\rm B}$ of the experimentally detected $dI/dV$ peak. Given these parameter choices, the calculated AA LDOS peaks occur at $E=-121$, $-84$, and $-59\,\mathrm{meV}$, providing a good approximation of the experimentally observed peak positions at $-117$, $-90$, and $-51\,\mathrm{meV}$, respectively.

\section{Heavy fermions embedded in an itinerant continuum in small-angle TBG}

The inclusion of the experimentally detected lattice dilation into our continuum model provides a phenomenological framework that successfully captures the impact of electro-mechanical coupling on the low-energy LDOS. In the following, we further show that the localized resonance at the AA site and the dense continuum of bands can be directly mapped into the heavy fermion model basis,
\begin{equation}
 H(\bk)=\begin{pmatrix}
 H_{ff}(\bk)&V_{fc}(\bk)\\
 V_{fc}^{\dagger}(\bk)&H_{cc}(\bk)
 \end{pmatrix}.
 \label{eq:fc_blocks}
\end{equation}
Here $H_{ff}$ acts within the localized doublet, $H_{cc}$ acts in its orthogonal complement $P_c=1-P_f$, and $V_{fc}$ couples the two sectors. 
The complement is constructed using the orthogonalized-plane-wave (OPW) approach of Ref.~\cite{shiHeavyFermionRepresentation2022}. This transformation retains the full plane-wave space used in the calculation and leaves the spectrum unchanged. Setting $V_{fc}=0$ allows us to independently examine the two sectors of this model [right inset of Fig.~\ref{fig:fig3}(c)]. In this limit, this projection yields two $H_{ff}$ bands around $-88\,\mathrm{meV}$, effectively dispersionless on the resolved energy scale, and a
dispersive $H_{cc}$ background. At small twist angles, the increased separation between AA-centered $f$ orbitals of nearly unchanged spatial extent suppresses intercell hopping, reducing their bandwidths far below the $\sim0.5\,\mathrm{meV}$ scale reported at twist angles near the magic angle~\cite{shiHeavyFermionRepresentation2022}. However, restoring $V_{fc}$ redistributes the localized weight among the continuum bands, accompanied by a small spectral shift to approximately $-85\,\mathrm{meV}$. The resulting $f-$orbital and $c-$state projected DOS in the bare and hybridized case are shown in Fig.~\ref{fig:fig3}(g,~h).

\begin{figure}[h]
    \centering
    \includegraphics[width=\textwidth]{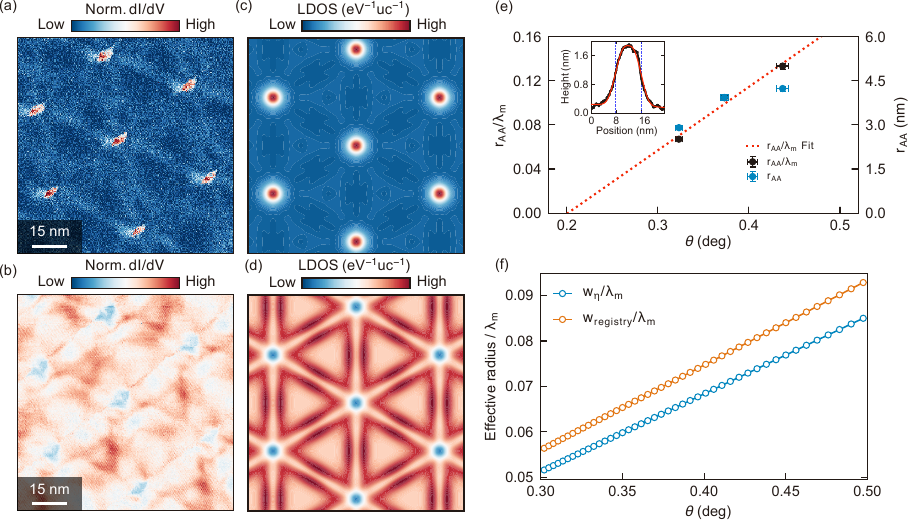}
    \caption{\justifying{\bf Heavy fermions embedded in an itinerant continuum in small-angle twisted bilayer graphene:} (a,~b) Differential conductance ($dI/dV$) maps recorded on small-angle TBG sample ($\theta \approx$  0.37\textsuperscript{$\circ$}) at the bias voltage of the $dI/dV$ peak ($V_{\rm B}=-85\,$mV) and near Fermi energy ($V_{\rm B}= -17\,$mV), respectively. (c,~d) Corresponding simulated local density of states (LDOS) maps at the energy of the AA site peak ($E-E_{F}= -85\,$meV) and near Fermi energy ($E-E_{F}= -4\,$meV), respectively. (e) Ratio of AA region radius $r_{AA}$ and the moiré wavelength $r_{AA} / \lambda_{m}$ as a function of twist angle $\theta$ extracted from STM topographic images of different devices. Inset: The value of $r_{AA}$ is determined by using super-Gaussian fitting (red line) of topographic line profiles along AA sites (black line). Details of this procedure are described in Supplementary Section~6.
    (f) Calculated equivalent-area radii of the dilation half-contrast region ($w_\eta$, blue) and the AA stacking-registry region ($w_{\rm registry}$, orange), normalized by the moir\'e period $\lambda_m$. Here the dilation profile is generated from the relaxed registry at each physical twist angle [see Supplementary Section~3].
    }
    \label{fig:fig4}
\end{figure}

This analysis reveals that the HF model, previously derived for describing the low-energy electronic structure of magic-angle TBG, naturally extends to smaller twist angles. We verify this understanding by considering spatially resolved maps of the differential tunnel conductance recorded on device A0 ($\theta \approx$  0.37\textsuperscript{$\circ$}). The strong localization of the $dI/dV$ peak at the AA site ($V_{B}=-85\,$mV) is successfully captured by the hybridized $f$-orbital state in our calculation [Fig.~\ref{fig:fig4}(a,~b)]. Meanwhile, our measurements near Fermi energy ($V_{B}=-17\,$mV) reveal that the narrow $dI/dV$ peaks [width $\approx10\,$mV] seen in our spectroscopic measurements [Fig.~\ref{fig:fig2}(h)] arise from spatially extended electronic states with triangular sub-structure at the AB sites [Fig.~\ref{fig:fig4}(c)]. This observation is closely mirrored by our model calculation [Fig.~\ref{fig:fig4}(d)], indicating that the electronic structure of small-angle TBG comprises strongly localized $f-$electrons embedded in itinerant $c-$electronic states.

We have also experimentally characterized the evolution of the HF picture at small twist angles by examining the spatial profile of the AA-localized features. To this end, we analyze the width of the AA site in STM topographies. We determine its effective spatial radius $r_{AA}$ by fitting a topographic line cut across AA sites using a Super-Gaussian profile [Fig.~\ref{fig:fig4}(e), see Supplementary Sec.~6 for details]. Our analysis reveals that the ratio $r_{AA}/\lambda_m$ decreases approximately linearly as the twist angle is reduced. This observation is qualitatively reproduced by the calculated equivalent-area radii of the dilation half-contrast region and the AA stacking-registry region [Fig.~\ref{fig:fig4}(f)] and is consistent with the theoretical understanding that in this strongly reconstructed regime, the absolute area of an AA region is nearly angle independent, while its fraction of the expanding moir\'e cell steadily decreases and the AB/BA domains occupy most of the cell
\cite{kazmierczakStrainFieldsTwisted2021,Nguyen2021Localization}.
Each moir\'e cell contributes a fixed integrated $f$-state spectral weight of two states per spin and valley, whereas the broad $c$-state DOS retains a nearly angle-independent overall scale per unit area. The $c-$state spectral weight per cell within a fixed energy window therefore increases with the moir\'e-cell area [see Fig.~\ref{fig:fig5} for twist-angle dependent DOS].
Consequently, while the local confinement scale of the $f-$orbitals remains comparable to that near the first magic angle, as shown by the local-orbital calculations, the decreasing AA area fraction reduces the relative contribution of AA-localized states to the low-energy manifold, leaving them embedded in the growing continuum of itinerant $c-$states.

\section{Discussion}

Our study demonstrates that lattice relaxation effects in small-angle TBG ($\theta<1$°) manifests in
a strongly particle-hole breaking electronic structure, qualitatively distinct from that observed at the first magic angle ($\theta=1.05$°). Consequently, the minimal Bistritzer--MacDonald continuum model~\cite{bistritzerMoireBandsTwisted2011}---which provides a reasonable approximation near the first magic angle but predicts a series of higher-order magic angles at $\theta<1\,$°---becomes insufficient without considering effects of relative moiré lattice reconstruction and atomic lattice dilation in this regime, as shown in this work. Moreover, while initially constructed to describe the physics near the first magic angle, we show that the HF model naturally extends to the small-angle regime where the appearance of compact, localized $f$-states at the AA-sites is found to be independent of the specific twist angle in experiment and theory. 

Thus, small-angle TBG provides a versatile platform to examine correlation-driven physics between partially filled $f$-orbitals without the need for strict fine-tuning to specific magic angles. This regime presents a physically distinct phenomenology compared to systems with spectrally isolated flat bands. Near the first magic angle, the $f$-orbital is approximately spectrally separated from the $c$-electrons over most of the moir\'e Brillouin zone [{\em c.f.~}Fig.~\ref{fig:fig1}(b)], except near the $\Gamma$ point~\cite{shiHeavyFermionRepresentation2022,SongBernevig2022},
meaning correlation effects within the partially filled flat band are predominantly governed by on-site Coulomb interactions~\cite{xie2019spectroscopic, wong2020cascade, zondiner2020cascade,SongBernevig2022}. Similarly, the flat bands in homo-bilayers of twisted transition metal dichalcogenides (TMD) are spectrally isolated, with low-energy physics dominated by local and non-local Coulomb interactions~\cite{wu2018hubbard, tang2020simulation, Xu2024WannierMoTe2, MoralesDuran2022Nonlocal}. 

\begin{figure}[h]
    \centering
    \includegraphics[width=1\textwidth]{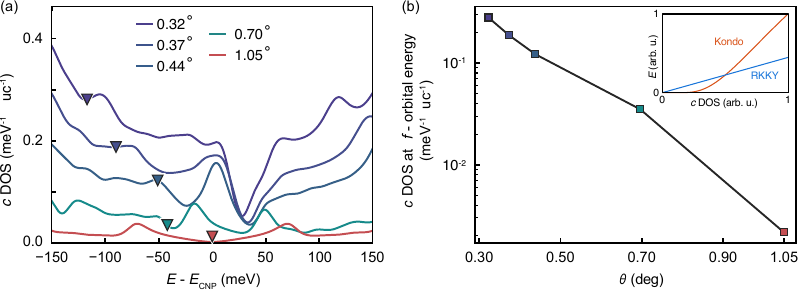}
    \caption{\justifying{\bf Toward Kondo lattice physics in small-angle twisted bilayer graphene:} (a) Energy ($E$) dependent density of states (DOS) per moir\'e cell of the itinerant electrons at different indicated twist angles. 
    Energies are measured relative to the charge-neutrality point $E_{\mathrm{CNP}}$, and the DOS is given per spin and valley. (b) DOS per moir\'e cell of itinerant $c$-states (extracted from panel a) at the energy of the $dI/dV$ peak at the AA site as a function of the twist angle $\theta$. The spectral position of the $dI/dV$ peaks [see Fig.~\ref{fig:fig2}(i)] is indicated by color-coded markers in panel a. Inset: Shown is the schematic dependence of the Kondo (red line) and RKKY (blue line) energy ($E$) scale as a function of the itinerant density of states.}
    \label{fig:fig5}
\end{figure}

Conversely, in small-angle TBG, flexo-electric coupling shifts the $f$-orbital to lower energies where they coexist with the itinerant $c-$states, as revealed in our study. The analysis of the calculated bare (not hybridized with $f-$orbitals) $c-$state DOS spectra at different twist angles [Fig.~\ref{fig:fig5}(a)] reveals that the DOS of the itinerant electronic states at the energy of the $f-$orbital scales approximately exponentially with decreasing twist angle [Fig.~\ref{fig:fig5}(b), also see Supplementary Section~5.D]. The itinerant background can effectively screen Coulomb interactions~\cite{Pizarro2019Screening, Barrier2026Screening, gao2026double}, leading to qualitatively different expected low-energy ground states than near the magic angle. Hence, the electronic structure of small-angle TBG bears similarity to the Kondo lattice in TMD hetero-bilayers~\cite{zhao2023gate}, however, with significantly enhanced $c-$DOS and larger $f-$orbital degeneracy (eightfold). Although a microscopic theoretical framework for the electro-mechanical coupling in TBG remains to be established, our study showcases that flexo-electric responses can realize novel electronic landscapes in moiré lattice materials beyond the single-particle picture. 

The interacting Kondo lattice problem of $f-$orbitals embedded in a $c-$continuum has been extensively studied in bulk heavy fermion compounds~\cite{lohneysen2007fermi, gegenwart2008quantum, si2010heavy} and more recently in TMD hetero-bilayers~\cite{zhao2024emergence}; however, revealing the physics of the quantum critical point separating Kondo-screened and long-range ordered phases has remained challenging. While the Ruderman-Kittel-Kasuya-Yosida interaction—which favors long-range magnetic order—depends linearly on the DOS at the Fermi energy, the Kondo energy scale depends exponentially on it [inset Fig.~\ref{fig:fig5}(b)]. Looking ahead, by utilizing the twist angle and electrostatic gating as tuning knobs for the $c-$DOS and $f-$orbital occupation (forming an effective $SU(8)$ space by spin, valley, and orbital-degeneracy), small-angle TBG could thus provide a model testbed for studying emergent ground states within the large$-N$ model~\cite{hu2023symmetric, chou2023kondo}.


\clearpage
\section{References}
\bibliography{references.bib}

\section{Acknowledgments}

We appreciate valuable discussions with Jonah Herzog-Arbeitman. This work was primarily supported by the Hong Kong RGC (Grant Nos. 26304221, 16302422, 16302624, 16304525, and C6033-22G) and the Croucher Foundation (Grant No. CIA22SC02) awarded to B.J. B.J. and X.D acknowledge support through the Area of Excellence Scheme by the Hong Kong RGC (Grant No. AoE/P-604/25-R).

\clearpage
\section{Methods}

\subsection{Device fabrication}
All devices were fabricated using dry-transfer method. Graphene and h-BN (thickness 15\,nm-30\,nm) flakes were mechanically exfoliated onto a 285\,nm SiO\textsubscript{2}/Si wafer using Scotch tape. The monolayer graphene was cut into half using an atomic-force-microscopy tip. Then, the TBG/h-BN heterostructure was picked up by a polyvinyl alcohol (PVA)/polycarbonate (PC)/polydimethylsiloxane (PDMS) stamp. The first half of the graphene was picked up by the stamp. The other part of the graphene (on the SiO\textsubscript{2}/Si wafer) was rotated by a target twist angle relative to the first half of the graphene and then picked up by the stamp. Subsequently, the h-BN flake was picked up and finally the TBG/h-BN heterostructure was released onto a SiO\textsubscript{2}/Si chip with prepatterned Cr/Au(5nm/50nm) electrodes. The polymer on the devices was then removed in subsequent organic solvent cleaning steps using N-methyl-2-pyrrolidone (NMP), deionized water, and isopropanol. The TBG surface was further cleaned with atomic-force-microscopy tip cleaning. Finally, devices were annealed in ultrahigh vacuum at 170$^{\circ}$C for 12\,h before being transferred to the STM.

\subsection{STM measurements}
STM measurements were performed using a home-built STM with a base pressure of $\approx1.4 \times 10^{-10}$\,mbar and base temperature of $T=4.5$\,K. The STM tip was prepared on a Cu(111) surface by field emission and controlled indention and calibrated against the Cu(111) Shockley surface state before each set of measurements. The differential conductance ($dI/dV$) spectra and maps were captured by a standard lock-in technique with a small bias modulation $V_{ac}$ ranging from 0.5 mV to 3 mV at a frequency of $f=719$\,Hz. $dI/dV$ maps were recorded using the multi-pass mode, where the topographic height profile was first recorded in constant current mode at a suitable sample bias voltage and then the position-dependent $dI/dV$ amplitude was recorded at a chosen sample bias voltage $V_{B}$ in a set of multiple passes along the same topographic line.

\clearpage
\section{Figures}

\clearpage

\clearpage

\clearpage

\clearpage

\end{document}